\documentclass[10pt]{article}
\usepackage{graphicx}
\usepackage{heppennames2}
\usepackage{lhchiggs}
\usepackage{cernunits}
\allowdisplaybreaks
\newcommand\pubdate{\today}

\def\csumb{Dipartimento di Fisica Teorica, Universit\`a di Torino, Italy\\
           INFN, Sezione di Torino, Italy}
\def\support{\footnote{Work supported by MIUR under contract
    2001023713$\_$006 and by Compagnia di San Paolo under contract ORTO11TPXK.}}
\def\Title#1{\begin{center} {\Large\bf #1 } \end{center}}

\def\Author#1{\begin{center}{ \sc #1} \end{center}}
\def\Address#1{\begin{center}{ \it #1} \end{center}}

\newcommand\pubblock{\rightline{\begin{tabular}{l} \\
         \pubdate\\  \end{tabular}}}
\newenvironment{Abstract}{\begin{quotation}  }{\end{quotation}}

\def\Acknowledgments{\bigskip  \bigskip \begin{center}
          \large\bf Acknowledgments\end{center}}
\def\email#1{\footnote{#1}}
\makeatletter
\def\section{\@startsection{section}{0}{\z@}{5.5ex plus .5ex minus
 1.5ex}{2.3ex plus .2ex}{\large\bf}}
\def\subsection{\@startsection{subsection}{1}{\z@}{3.5ex plus .5ex minus
 1.5ex}{1.3ex plus .2ex}{\normalsize\bf}}
\def\subsubsection{\@startsection{subsubsection}{2}{\z@}{-3.5ex plus
-1ex minus  -.2ex}{2.3ex plus .2ex}{\normalsize\sl}}

\renewcommand{\@makecaption}[2]{%
   \vskip 10pt
   \setbox\@tempboxa\hbox{\small #1: #2}
   \ifdim \wd\@tempboxa >\hsize     
       \small #1: #2\par          
     \else                        
       \hbox to\hsize{\hfil\box\@tempboxa\hfil}
   \fi}
 \def\citenum#1{{\def\@cite##1##2{##1}\cite{#1}}}
\def\citea#1{\@cite{#1}{}}
\newcount\@tempcntc
\def\@citex[#1]#2{\if@filesw\immediate\write\@auxout{\string\citation{#2}}\fi
  \@tempcnta\z@\@tempcntb\m@ne\def\@citea{}\@cite{\@for\@citeb:=#2\do
    {\@ifundefined
       {b@\@citeb}{\@citeo\@tempcntb\m@ne\@citea\def\@citea{,}{\bf }\@warning
       {Citation `\@citeb' on page \thepage \space undefined}}%
    {\setbox\z@\hbox{\global\@tempcntc0\csname b@\@citeb\endcsname\relax}%
     \ifnum\@tempcntc=\z@ \@citeo\@tempcntb\m@ne
       \@citea\def\@citea{,}\hbox{\csname b@\@citeb\endcsname}%
     \else
      \advance\@tempcntb\@ne
      \ifnum\@tempcntb=\@tempcntc
      \else\advance\@tempcntb\m@ne\@citeo
      \@tempcnta\@tempcntc\@tempcntb\@tempcntc\fi\fi}}\@citeo}{#1}}
\def\@citeo{\ifnum\@tempcnta>\@tempcntb\else\@citea\def\@citea{,}%
  \ifnum\@tempcnta=\@tempcntb\the\@tempcnta\else
  {\advance\@tempcnta\@ne\ifnum\@tempcnta=\@tempcntb \else\def\@citea{--}\fi
    \advance\@tempcnta\m@ne\the\@tempcnta\@citea\the\@tempcntb}\fi\fi}
\makeatother
\DeclareRobustCommand{\PA}{\HepParticle{A}{}{}\Xspace}
\DeclareRobustCommand{\PV}{\HepParticle{V}{}{}\Xspace}
\DeclareRobustCommand{\PX}{\HepParticle{X}{}{}\Xspace}
\DeclareRobustCommand{\Pf}{\HepParticle{f}{}{}\Xspace}

\newcommand{\myLO}{\rm{\scriptscriptstyle{LO}}}

\newcommand{\myNNLO}{\rm{\scriptscriptstyle{NNLO}}}

\newcommand{\ssA}{{\mathrm{A}}}
\newcommand{\ssB}{{\mathrm{B}}}

\newcommand{\ssF}{{\mathrm{F}}}
\newcommand{\ssR}{{\mathrm{R}}}
\newcommand{\ssD}{{\mathrm{D}}}

\newcommand{\ssS}{{\mathrm{S}}}
\newcommand{\ssI}{{\mathrm{I}}}

\newcommand{\ssM}{{\mathrm{M}}}

\newcommand{\ssZ}{{\mathrm{Z}}}
\newcommand{\ssZZ}{{\mathrm{ZZ}}}
\newcommand{\bqas}{\begin{eqnarray*}}
\newcommand{\eqas}{\end{eqnarray*}}

\newcommand{\lpar}{\left(}                            
\newcommand{\rpar}{\right)}

\newcommand{\bq}{\begin{equation}}                    
\newcommand{\eq}{\end{equation}}
\newcommand{\bqa}{\arraycolsep 0.14em\begin{eqnarray}}
\newcommand{\eqa}{\end{eqnarray}}
\newcommand{\ba}[1]{\begin{array}{#1}}
\newcommand{\ea}{\end{array}}
\newcommand{\ben}{\begin{enumerate}}
\newcommand{\een}{\end{enumerate}}
\newcommand{\bei}{\begin{itemize}}
\newcommand{\eei}{\end{itemize}}
\newcommand{\eqn}[1]{Eq.(\ref{#1})}

\newcommand{\bmid}{\Bigr|}

\newcommand{\ord}[1]{{\cal O}\lpar#1\rpar}

\newcommand{\Bref}[1]{Ref.~\cite{#1}}
\newcommand{\Brefs}[1]{Refs.~\cite{#1}}

\newcommand{\eg}{e.g.\xspace}
\newcommand{\ie}{i.e.\xspace}
\newcommand{\etc}{etc.\@\xspace}

\newcommand{\cf}{\emph{cf.\xspace}}

\newcommand{\mBh}{\mathswitch {{\overline M}_{\PH}}}
\newcommand{\mBhs}{\mathswitch {{\overline M}^2_{\PH}}}

\newcommand{\cph}{\mathswitch {s_{\PH}}}

\newcommand{\muh}{\mathswitch {\mu_{\PH}}}
\newcommand{\muph}{\mathswitch {\mu'_{\PH}}}
\newcommand{\muhs}{\mathswitch {\mu^2_{\PH}}}
\newcommand{\gh}{\mathswitch {\gamma_{\PH}}}

\newcommand{\GOL}{\mathswitch {\overline\Gamma}}

\newcommand{\muR}{\mathswitch {\mu_{\ssR}}}
\newcommand{\muF}{\mathswitch {\mu_{\ssF}}}

\newcommand{\myprod}{{\mbox{\scriptsize prod}}}
\newcommand{\prop}{{\mbox{\scriptsize prop}}}

\newcommand{\ac}{{\mbox{\scriptsize all}}}
\newcommand{\rest}{{\mbox{\scriptsize rest}}}
\newcommand{\eff}{{\mbox{\scriptsize eff}}}

\providecommand\HTO{{\sc HTO}}

\begin{document}
\begin{titlepage}
\pubblock
%
\vfill
\def\thefootnote{\fnsymbol{footnote}}
\Title{Higgs Interference Effects in $\Pg \Pg \to \PZ\PZ$\\[0.1cm]
and their Uncertainty\support}
\vfill
\Author{Giampiero Passarino\email{giampiero@to.infn.it}}
\Address{\csumb}
\vfill
\vfill
\begin{Abstract}
\noindent 
Interference between the Standard Model Higgs boson and continuum contributions in 
$\Pg\Pg \to \PZ\PZ$ is considered in the heavy-mass scenario. Results are available at 
leading order for the background (the $\Pg\Pg \to \PZ\PZ$ box diagrams). It is discussed how 
to combine the result with the next-to-next-to-leading order Higgs production cross-section 
and a proposal for estimating the associated theoretical uncertainty is presented. 
\end{Abstract}
\vfill
\begin{center}
\end{center}
\end{titlepage}
\def\thefootnote{\arabic{footnote}}
\setcounter{footnote}{0}
\small
\thispagestyle{empty}
%
\normalsize
\clearpage
\setcounter{page}{1}
\section{Introduction \label{intro}}
At the beginning of $2011$ the status of the inclusive cross-section for Higgs-boson production 
in gluon fusion was summarized~\cite{Dittmaier:2011ti}. Corrections arising from 
higher-order QCD, electroweak effects, as well as contributions beyond the commonly-used 
effective theory approximation were analyzed. Uncertainties arising from missing terms in the 
perturbative expansion, as well as imprecise knowledge of parton distribution functions, 
were estimated to range from approximately $15{-}20\%$, with the precise value depending on 
the Higgs boson mass. For an updated study we refer to \Bref{Dittmaier:2012vm}.

Recently the problem of going beyond the zero-width approximation has received 
new boost from the work of \Brefs{Anastasiou:2011pi,Anastasiou:2012hx} and of
\Bref{Goria:2011wa} which implemented the complex-pole-scheme with an estimate of
the residual theoretical uncertainty (see also the work of \Bref{Kauer:2012hd}).
Here we only recall that the complex pole describing an unstable particle is conventionally 
parametrized as
\bq
s_i = \mu^2_i - i\,\mu_i\,\gamma_i,
\label{CPpar}
\eq
with $i= \PW, \PZ, \PH$ \etc (see \Bref{Goria:2011wa,Passarino:2010qk,Actis:2006rc}).

In the current experimental analysis there are additional sources of uncertainty, \eg background 
and Higgs interference effects~\cite{ATLAS:2012ac,pippo-atlas,Chatrchyan:2012ty,pluto-atlas,Chatrchyan:2012ft}. As a matter of fact, this interference is partly available and should not be included 
as a theoretical uncertainty; for a discussion and results we refer to 
\Brefs{Campbell:2011cu,Kauer:2012ma,Binoth:2006mf,Accomando:2007xc}.
In particular, from \Brefs{Campbell:2011cu} we see that at $\muh= 600\UGeV$ (the highest value
reported) the effect is about $+40\%$ in the window $\zeta= 440{-}560\UGeV$, where $\zeta$ is
the Higgs virtuality; the effect is practically zero at the peak and reaches $-50\%$ after 
$\zeta= 680\UGeV$ (no cuts applied). 
For the total cross-section in $\Pg\Pg \to \Pl\PGn\Pl'\PGn'$ at $\muh = 600\UGeV$ the
effect of including the interference is already $+34\%$ and rapidly increasing with
$\muh$.

We stress that setting limits without including the effects of the interference induces 
large variations in rate and shape that will propagate through to all distributions. Therefore, 
any attempt to analyze kinematic distributions which are far from the Standard Model (SM) shape 
may result in misleading limits.

The importance of a complete understanding of the shape of the background from non-resonant
diagrams has been emphasised in \Brefs{Baur:1990af,Baur:1990mr}. It was shown in this work 
that a heavy Higgs boson with mass larger than $800\UGeV$ does not lead to a pronounced peak 
structure in the lineshape and predictions for the non-resonant background must therefore be 
as accurate as possible in order to discriminate a heavy Higgs boson from a light one.

In the current experimental analysis for heavy Higgs searches, a theoretical uncertainty 
of $150\,\muh^3[\%]$ ($\muh$ in \UTeV) has been used for conservative 
estimate\footnote{https://twiki.cern.ch/twiki/bin/view/LHCPhysics/HeavyHiggs}.
For a Higgs boson of $700\UGeV$ this amounts to $\pm 51\%$.

One might wonder why considering a Standard Model (SM) Higgs boson in such a high-mass range. 
There are classic constraints on the Higgs boson mass coming from unitarity, triviality and
vacuum stability, precision electroweak data and absence of fine-tuning~\cite{Ellis:2009tp}. 
The situation is different if we consider extensions of the SM: in the THD model, even if the
SM-like Higgs boson is found to be light ($< 140\UGeV$), there is a possible range of mass
splitting in the heavy Higgs boson. In general, for a given Higgs boson mass, the magnitude
of the mass splittings among different heavy scalar bosons can be determined to satisfy the
electroweak precision data, see \Bref{Kanemura:2011sj}.

This paper is organized as follows. In \refS{Sect_summary} we discuss results available
in the literature. In \refS{Sect_num} we present and discuss numerical results.
Conclusions are presented in \refS{Sect_conc}. 
\section{Summary of available results \label{Sect_summary}}
There are serious problems in including the signal/background interference in gluon-gluon fusion
and very few examples of theoretical predictions, \eg interference has been computed for the
di-photon signal in \Bref{Dixon:2003yb}.
Let us concentrate on the process $\Pg \Pg \to \PZ \PZ$, the whole cross-section can be written 
as follows (here, for simplicity, we neglect folding the partonic process with parton
distribution functions):
\bq
\sigma_{\Pg \Pg \to \PZ \PZ} = 
\sigma_{\Pg \Pg \to \PZ \PZ}(S) + 
\sigma_{\Pg \Pg \to \PZ \PZ}(I) + 
\sigma_{\Pg \Pg \to \PZ \PZ}(B),  
\eq
where $S, B$ and $I$ stand for signal ($\Pg\Pg \to \PH \to \PZ\PZ$), background
($\Pg\Pg \to \PZ\PZ$, \ie gluon initiated box contribution for all three quark doublets)
and interference; the signal can be written as
\bq
\sigma_{\Pg \Pg \to \PZ \PZ}(S) =
\sigma_{\Pg \Pg \to \PH \to \PZ \PZ}(\zeta) = \frac{1}{\pi}\,
\sigma_{\Pg \Pg \to \PH}\,\frac{\zeta^2}{\bmid \zeta - \cph\bmid^2}\,
\frac{\Gamma_{\PH \to \PZ\PZ}}{\sqrt{\zeta}},
\eq
\label{signal}
where we have introduced $\cph$, the Higgs boson complex 
pole~\cite{Passarino:2010qk,Goria:2011wa}; furthermore, $\zeta$ is the Higgs boson virtuality.
So far, most if not all theoretical prediction have been devoted to compute the signal with
the highest possible precision, next-to-leading order (NLO) and beyond. Therefore, in 
\eqn{signal}, we have
\bei

\item the production cross-section, $\sigma_{\Pg \Pg \to \PH}$, with next-to-next-to-leading 
logarithmic resummation (NNLL)~\cite{Catani:2003zt}, \ie with NNLL ${+}$ NNLO ${+}$ EW ${+}$ 
bottom quark contribution up to NLO ${+}$ NLL (three loop level plus resummation), see 
\Bref{Dittmaier:2012vm}

\item the partial decay width of an off-shell Higgs boson of virtuality $\zeta$, 
$\Gamma_{\PH \to \PZ\PZ}$, which is known at NLO (one-loop) with leading NNLO effects in the 
limit of large Higgs boson mass, see \Bref{Bredenstein:2007ec,Prophecy4f}.

\eei
However the background (continuum $\Pg \Pg \to \PZ \PZ$) and the interference are only known
at leading order (LO, one-loop)~\cite{Glover:1988rg}. Here we face two problems, a missing 
NLO calculation of the background (two-loop) and the NLO or NNLO signal at the amplitude level, 
without which there is no way to improve upon the present LO calculation.
For low values of the Higgs boson mass the interference arises primarily from the imaginary part 
of the continuum background interfering with the real part of the signal.  
For $\muh < 2\,m_{\PQt}$, where we can use the effective theory (\ie large-$\,m_{\PQt}$ limit), 
it would be relatively easy to get the signal amplitude; however, what might be the tougher 
part is implementing it in the same program that contains the background. Above the 
$\PAQt\PQt\,$-threshold, getting the signal amplitude becomes more difficult. We know that 
the effective theory misses imaginary parts in this region, and it is not clear how one would 
trust a calculation for the interference using it.  

Of course, putting in the $\ord{\alphas^n}$ corrections to the signal without the background can 
only be considered an approximation to the interference correction. It is difficult
to believe that background NLO calculation will be done in a foreseeable future since
not all basic master integral for the two-loop contribution are known, at least analytically.

We can also summarize the basic features of the LO interference, especially for high
value of the Higgs boson mass where the contribution from the imaginary part of the signal is
not negligible.
The main issue is on unitarity cancellations at high energy. Of course, the behavior of both 
LO amplitudes (signal and background) for $M_{\ssZZ} \to \infty$ is known and simple
and any correct treatment of perturbation theory (no mixing of different orders) will 
respect the unitarity cancellations. Since the Higgs boson decays almost completely into 
longitudinal $\PZ$s, for $M_{\ssZZ} \to \infty$ we have~\cite{Glover:1988rg} (for a single 
quark $q$)
\bq
A_{\ssS} \sim \frac{M^2_{\ssZZ} m^2_q}{2 M^2_{\PZ}}\,\Delta_{\PH}\,\ln^2\frac{M^2_{\ssZZ}}{m^2_q}
\qquad
A_{\ssB} \sim -\,\frac{m^2_q}{2 M^2_{\PZ}}\,\ln^2\frac{M^2_{\ssZZ}}{m^2_q},
\label{asym}
\eq
where $\Delta_{\PH}$ is the Higgs propagator, showing cancellation in the limit 
($M^2_{\ssZZ}\,\Delta_{\PH} \to 1$). However, the behavior for $M^2_{\ssZZ} \to \infty$ 
(unitarity) should not/cannot be used to simulate the interference for $M_{\ssZZ} < \muh$.
The only relevant message to be derived here is that unitarity requires the interference
to be destructive at large values of $M_{\ssZZ}$. The explicit LO calculation also shows that the
interference is constructive below the Higgs peak.
The higher-order correction in gluon-gluon fusion~\cite{Dawson:1991zj,Djouadi:1991tka,Spira:1995rr,Kramer:1996iq,Harlander:2001is,Anastasiou:2002yz,Ravindran:2002dc,Catani:2003zt}
have shown a huge $K\,$-factor (for updated cross-sextions at $8\UTeV$ see 
\Bref{deFlorian:2012yg}) 
\bq
K = \frac{\sigma_{\myprod}^{\myNNLO}}{\sigma_{\myprod}^{\myLO}},
\qquad
\sigma_{\myprod} = \sigma_{\Pg\Pg \to \PH}.
\eq
A potential worry, already addressed in \Bref{Campbell:2011cu}, is: should we simply use the 
full LO calculation or should we try to effectively include the large (factor two) $K\,$-factor
to have effective NNLO observables? There are different opinions since interference effects 
may be as large or larger than NNLO corrections to the signal. Therefore, it is important to 
quantify both effects. So far, two options have been introduced to account for the increase in 
the signal (\cf \Bref{Ciar,Frix}). Let us consider any distribution $D$, \ie
\bq
D = \frac{d\sigma}{d M_{\ssZZ}} \quad \mbox{or} \quad \frac{d\sigma}{d \pT^{\ssZ}} \quad
\mbox{\etc}
\eq
where $M_{\ssZZ}$ is the invariant mass of the $\PZ\PZ\,$-pair and $\pT^{\ssZ}$ is the
transverse momentum. Two possible options are:
\begin{itemize}
\item {\bf{additive}} where one computes
\bq
D^{\myNNLO}_{\eff} = D^{\myNNLO}(S) + D^{\myLO}(I) + D^{\myLO}(B)
\label{Aopt}
\eq
\item {\bf{multiplicative}} where one computes
\bq
D^{\myNNLO}_{\eff} = K_{\ssD}\,\bigl[ D^{\myLO}(S) + D^{\myLO}(I) \bigr] + D^{\myLO}(B),
\qquad
K_{\ssD} = \frac{D^{\myNNLO}(S)}{D^{\myLO}(S)},
\label{Mopt}
\eq
where $K_{\ssD}$ is the differential $K\,$-factor for the distribution.
\end{itemize}
In both cases the NNLO corrections include the NLO electroweak part, for 
production~\cite{Actis:2008ug} and decay~\cite{Prophecy4f}. It is worth noting that the 
differential $K\,$-factor for the $\PZ\PZ\,$-invariant mass distribution is a slowly increasing 
function of $M_{\ssZZ}$, going (\eg for $\muh= 700\UGeV$) from $2.04$ at $M_{\ssZZ} = 210\UGeV$ to
$2.52$ at $M_{\ssZZ} = 1\UTeV$.

The two options, as well as intermediate ones, suffer from an obvious problem: they are spoiling 
the unitarity cancellation between signal and background for $M_{\ssZZ} \to \infty$, 
breakdown which is described in details in \Bref{Glover:1988rg}. Therefore, our partial conclusion 
is that any option showing an early onset of unitarity violation should not be used for too high 
values of the $\PZ\PZ\,$-invariant mass.   

Therefore, our first prescription in proposing an effective higher-order interference will be
to limit the risk of overestimation of the signal by applying the recipe only in some
restricted interval of the $\PZ\PZ\,$-invariant mass, \eg $[\muh - \gh\,,\,\muh + \gh]$.
This is especially true for high values of $\muh$ where the width is large.

Explicit calculations show that the {\em multiplicative} option is better suited for regions with
destructive interference while the {\em additive} option can be used in regions where the effect 
of the interference is positive, \ie we still miss higher orders from the background amplitude
but do not spoil cancellations between signal and background.

Actually, there is an intermediate options that is based on the following observation:
higher-order corrections to the signal are made of several terms (see \Bref{Spira:1995rr} for 
a definition of $\Delta\sigma$),
\bq
\sigma^{\myprod} = 
\sum_{i,j}\,\int \hbox{PDF}\,\otimes\,\sigma^{\myprod}_{i j \to \ac} = 
\sum_{i,j}\,\int_{z_0}^1 dz \int_z^1 \frac{dv}{v}\,{\cal L}_{ij}(v)
\sigma^\prop_{ij \to \ac}(\zeta, \kappa, \muR, \muF),
\label{PDFprod_1}
\eq
where the sum is over incident partons; furthermore $\zeta = z\,s$ is the Higgs virtuality, 
$z_0$ is a lower bound on the invariant mass of the $\PH$ decay products, $\kappa= v\,s$ is the 
invariant mass of the incoming partons ($i, j$) and the luminosity is defined by
\bq
{\cal L}_{ij}(v) = \int_v^1 \frac{dx}{x}\,
f_i\lpar x,\muF\rpar\,f_j\lpar \frac{v}{x},\muF\rpar.
\label{PDFprod_2}
\eq
The partonic cross-section is defined by
\bq
\sum_{ij}\,\sigma_{ij \to \PH  + \PX}(\zeta, \kappa, \muR, \muF) = 
\sigma_{\Pg\Pg \to \PH}\,\delta\lpar 1 - \frac{z}{v}\rpar + \frac{s}{\kappa}\,\lpar
\Delta\sigma_{\Pg\Pg \to \PH\Pg} + \Delta\sigma_{\PQq\Pg \to \PH\PQq} + 
\Delta\sigma_{\PAQq\PQq \to \PH\Pg} + \mbox{NNLO}\rpar.  
\label{PDFprod_4}
\eq
From this point of view it seems more convenient to define
\bq
K_{\ssD} = K_{\ssD}^{\Pg\Pg} + K_{\ssD}^{\rest},
\qquad
K_{\ssD}^{\Pg\Pg} = \frac{D^{\myNNLO}\lpar \Pg\Pg \to \PH(\Pg) \to \PZ\PZ(\Pg)\rpar}
{D^{\myLO}\lpar \Pg\Pg \to \PH \to \PZ\PZ\rpar}
\eq
and to introduce a third option
\begin{itemize}
\item {\bf{intermediate}}
\bq
D^{\myNNLO}_{\eff} = K_{\ssD}\,D^{\myLO}(S) + \lpar K_{\ssD}^{\Pg\Pg}\rpar^{1/2}\,D^{\myLO}(I) 
+ D^{\myLO}(B)
\label{Iopt}
\eq
which, in our opinion, better simulates the inclusion of $K\,$-factors at the level of
amplitudes (although we are still missing corrections to the continuum amplitude).

\end{itemize}

Alternatively one could consider a different approach when $M_{\ssZZ} >> \muh$; it is based
on \eqn{asym} and on the work of \Bref{Seymour:1995qg} and amounts to neglect the
background (where NLO corrections are not available) while modifying the Higgs propagator,
\bq
\frac{1}{M^2_{\ssZZ} - \cph} =
\Bigl( 1 + i\,\frac{{\GOL}_{\PH}}{\mBh}\Bigr)\,
\Bigr( M^2_{\ssZZ} - \mBhs + i\,\frac{{\GOL}_{\PH}}{\mBh}\,M^2_{\ssZZ} \Bigr)^{-1}
\quad \to \quad
\frac{\mBhs}{M^2_{\ssZZ}}\,
\Bigr( M^2_{\ssZZ} - \mBhs + i\,\frac{{\GOL}_{\PH}}{\mBh}\,M^2_{\ssZZ} \Bigr)^{-1}
\label{barid}
\eq
where we have introduced mass and width in the Bar-scheme~\cite{Goria:2011wa} according to
\bq
\mBhs = \muhs + \gamma^2_{\PH} 
\qquad
\muh\,{\GOL}_{\PH} = \mBh\,\gh.
\label{Bars}
\eq
This recipe, \eqn{barid}, should be used only for $M_{\ssZZ} >> \muh$ and should not be extended 
below the resonant peak.

In the following Section we present numerical results in the high Higgs-mass region.
\section{Numerical results \label{Sect_num}}
In the following we will present numerical results obtained with the program 
\HTO{} (G.~Passarino, unpublished) that allows for the study of the Higgs--boson-lineshape, in
gluon-gluon fusion (ggF), using complex poles. \HTO{} is a FORTRAN $95$ program that contains a 
translation of the subroutine {\tt HIGGSNNLO} written by M.~Grazzini for computing the total 
(on-shell) cross-section for Higgs-boson production (in ggF) at NLO and 
NNLO~\cite{deFlorian:2009hc,Grazzini:2008tf,Catani:2001ic} and a translation of the
program {\tt ggzz} by E.W~Glover and J.J.~van der Bij for computing the LO interference in
$\Pg\Pg \to \PZ\PZ$.

All results in this paper refer to $\sqrt{s} = 8\UTeV$ and are based on the MSTW2008 PDF 
sets~\cite{Martin:2009iq}. They are implemented according to the OFFP - scheme, see
Eq.(45) of \Bref{Goria:2011wa}. Furthermore we use renormalization and factorization QCD scales
that evolve with the Higgs virtuality ($M_{\ssZZ}$).

In \refT{tab:HTO_1} we present the effect of the interference w.r.t. signal $\,+\,$ background
for the total cross-section. We select a leptonic final state (the branching ratio for
both $\PZ$ bosons to decay into $\Pe$ or $\PGm$ is $4.36\,10^{-3}$) and 
use $\pT^{\ssZ} > 0.25\,M_{\ssZZ}$ and $2\,M_{\PZ} < M_{\ssZZ} < 1\UTeV$. As is evident the 
strong cancellations between the constructive and destructive interference below and above the 
peak result in a small effect on the total cross-section.
\begin{table}[ht]
\begin{center}
\caption[]{\label{tab:HTO_1}{Interference effect for the total cross-section at LO and NNLO 
with multiplicative (M) option of \eqn{Mopt}, additive (A) option of \eqn{Aopt} and
intermediate option of \eqn{Iopt}. Here $\pT^{\ssZ} > 0.25\,M_{\ssZZ}$ and 
$2\,M_{\PZ} < M_{\ssZZ} < 1\UTeV$. In the table we show the percentage effect of the
interference, \ie interference/(signal + background).}}
\vspace{0.2cm}
\begin{tabular}{ccccc}
\hline 
$\muh$[GeV] & LO         & NNLO(A)    & NNLO(I)    & NNLO(M) \\
$400$       & $0.80[\%]$ & $0.64[\%]$ & $1.05[\%]$ & $1.65[\%]$ \\
$600$       & $0.98[\%]$ & $0.93[\%]$ & $1.57[\%]$ & $2.52[\%]$ \\
$800$       & $0.66[\%]$ & $0.63[\%]$ & $1.12[\%]$ & $1.84[\%]$ \\
\hline
\end{tabular}
\end{center}
\end{table}
However, the effect is drastically different on distributions. We present results 
for the $M_{\ssZZ}\,$-distribution (lineshape) with a cut $\pT^{\ssZ} > 0.25\,M_{\ssZZ}$, where 
$M_{\ssZZ}$ is the invariant mass of the $\PZ\PZ\,$-pair.

In \refF{fig:HTO_1} we show the lineshape for a Higgs mass of $600\UGeV$. The black line
gives the full $\Pg\Pg \to \PZ\PZ$ process at LO; the cyan line gives signal plus
background (LO) neglecting interference while the blue line includes both $\Pg\Pg$ and
$\PAQq\PQq$ initial states (LO). The red line gives the LO signal with different cuts
on the $\PZ$ transverse momentum. 

In \refF{fig:HTO_2} we present options for including higher-order effects. The black line is 
again full LO $\Pg\Pg \to \PZ\PZ$ result, the brown line gives the multiplicative option, the 
red line is the additive option while the blue line is the intermediate option. 
The cyan line gives signal plus background (LO) neglecting interference.

In \refF{fig:HTO_3} we show the same set of results as in \refF{fig:HTO_2} but for a Higgs boson 
mass of $700\UGeV$.

There are different options for showing the effect of the interference, \eg
\bq
R_{\eff} = \frac{D^{\myNNLO}_{\eff}}{D^{\myLO}(S) + D^{\myLO}(B)} - 1,
\qquad
D= \frac{d\sigma}{d M_{\ssZZ}}.
\label{perc}
\eq
However, our preferred way will be to use the following equation:
\bq
R'_{\eff} = \frac{K_{\ssD}\,D^{\myLO}(S) + 
\lpar K_{\ssD}^{\Pg\Pg}\rpar^{1/2}\,D^{\myLO}(I) + D^{\myLO}(B)}
{K_{\ssD}\,D^{\myLO}(S) + D^{\myLO}(B)} - 1,
\qquad
D= \frac{d\sigma}{d M_{\ssZZ}}.
\label{percp}
\eq
To summarize:
\bei
\item[] \eqn{percp} is our recipe for estimating the theoretical uncertainty in the effective
NNLO distribution: the intermediate option gives the {\em central value}, while the band
between the multiplicative and the additive options gives the uncertainty.
\eei
Note that the difference between the intermediate option and the median of the band is 
always small if not far away from the peak where, in any case, any option becomes questionable. 
The ratio $K_{\ssD}^{\Pg\Pg}/K_{\ssD}$ can be greater than one in some region, \eg for
$M_{\ssZZ} gtrsim 316 UGeV$, almost $\muh\,$-independent with a maximum of $1.024$ at 
$M_{\ssZZ} = 1 TeV$.

\begin{figure}
\vspace{-5cm}
\begin{minipage}{.9\textwidth}
\begin{center}
  \includegraphics[width=1.2\textwidth, bb = 0 0 595 842]{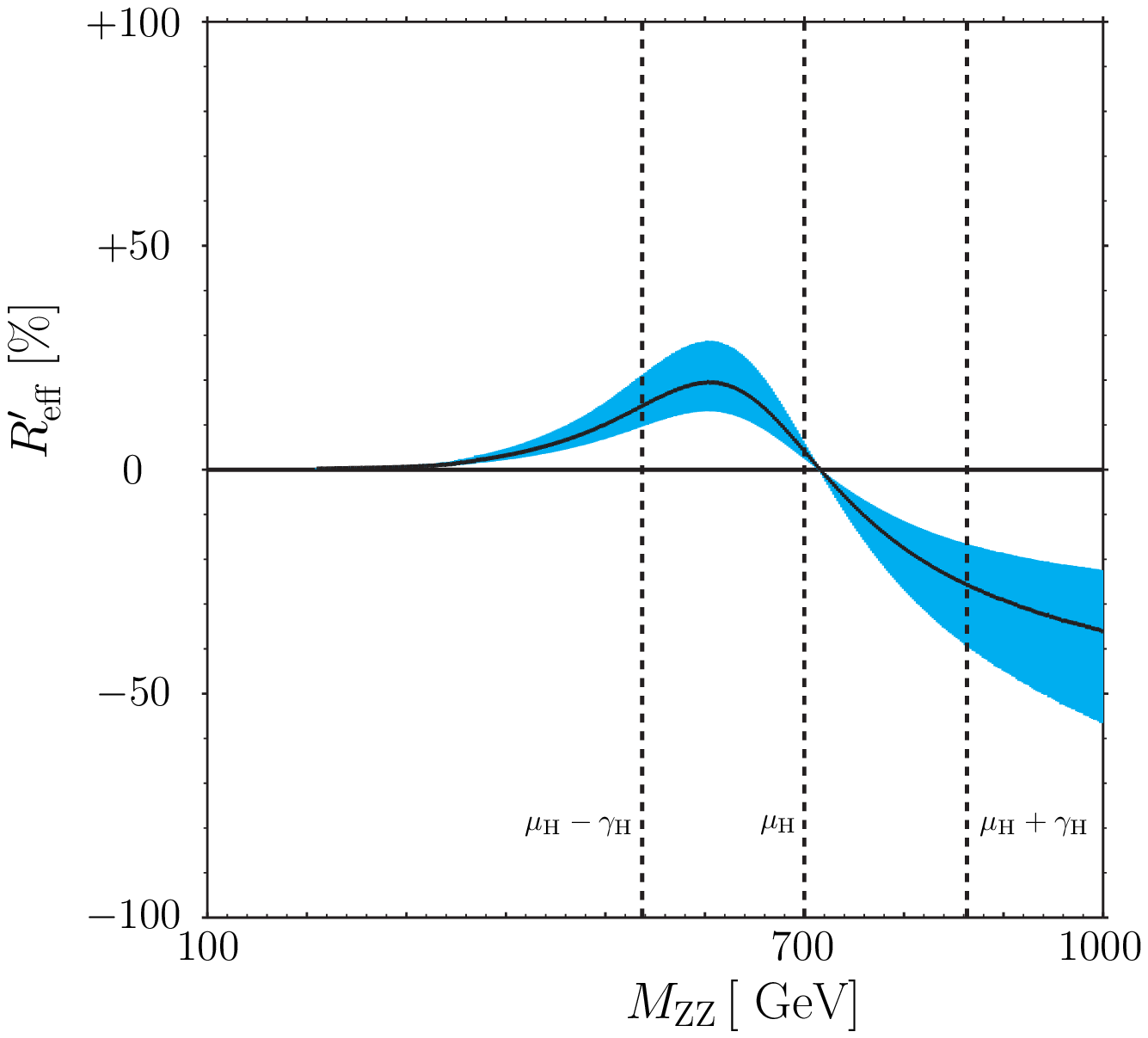}
  \vspace{-3.6cm}
  \caption{
Interference effects (see \eqn{percp}) in the $\PZ\PZ$ distribution for $\muh= 700\UGeV$.
The black line is the central value, the blue lines give the estimated theoretical uncertainty.}
\label{fig:HTO_5}
\end{center}
\end{minipage}
\end{figure}

\begin{figure}
\vspace{-5cm}
\begin{minipage}{.9\textwidth}
\begin{center}
  \includegraphics[width=1.2\textwidth, bb = 0 0 595 842]{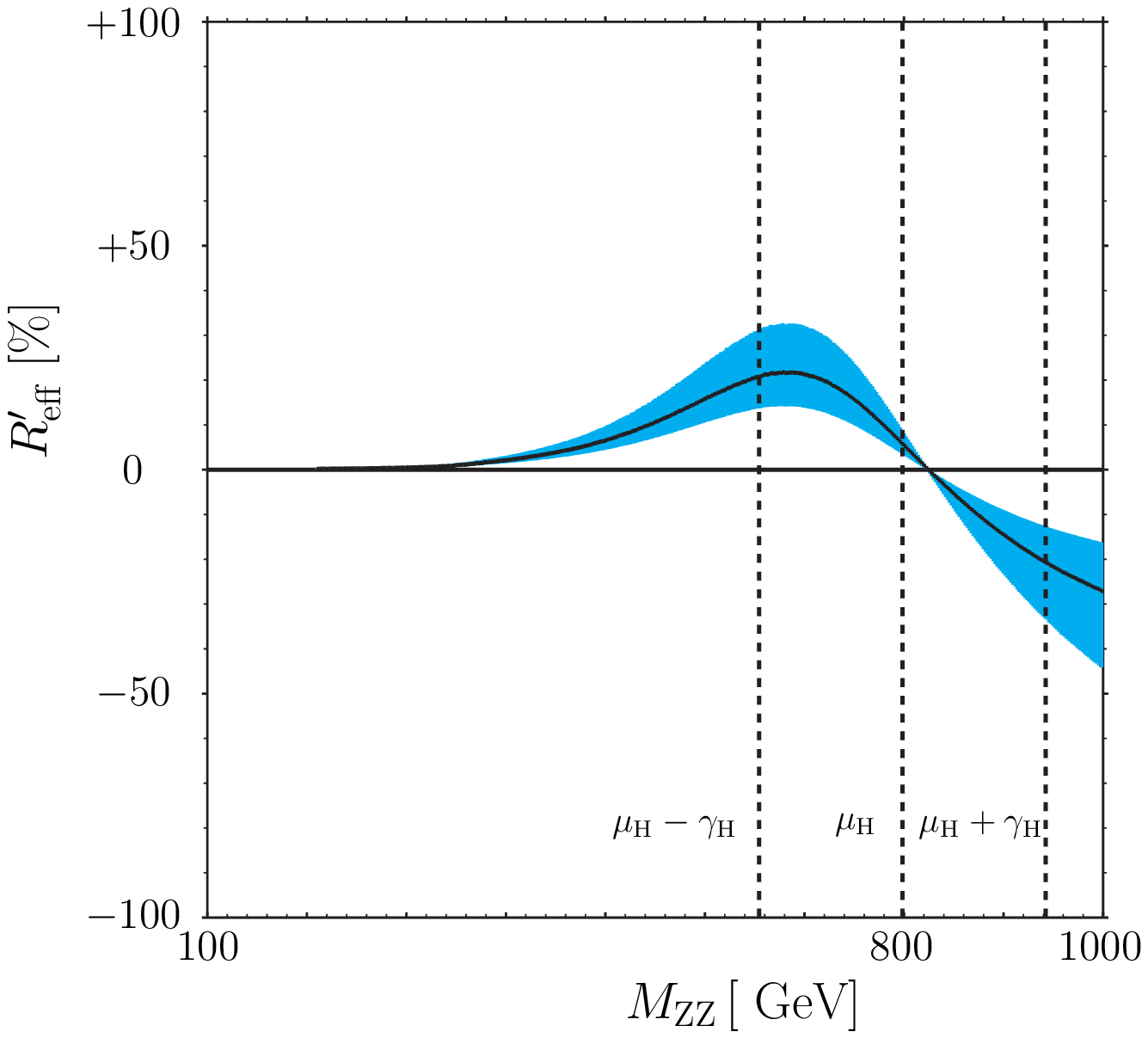}
  \vspace{-3.6cm}
  \caption{
Interference effects (see \eqn{percp}) in the $\PZ\PZ$ distribution for $\muh= 800\UGeV$.
The black line is the central value, the blue lines give the estimated theoretical uncertainty.}
\label{fig:HTO_6}
\end{center}
\end{minipage}
\end{figure}

In \refT{tab:HTO_2} we show the estimated theoretical uncertainty for the fractional
interference correction to the $700\UGeV$ resonance, $R'_{\eff}$ defined in \eqn{percp}.
The effect computed according to \eqn{percp} is very similar to the one obtained by considering
LO alone, shifted to the left and slightly less destructive for high values of $M_{\ssZZ}$.
In \refF{fig:HTO_5}(\refF{fig:HTO_6}) we present $R'_{\eff}$ (\eqn{percp}) for 
$\muh= 700\UGeV(800\UGeV)$ summarizing the percentage effect of interference in the effective 
NNLO theory. The black line gives the central value while the two blue lines represent the 
estimated theoretical uncertainty in including the NNLO $K\,$-factor. 

\begin{figure}
\vspace{-5cm}
\begin{minipage}{.9\textwidth}
\begin{center}
  \includegraphics[width=1.0\textwidth, bb = 0 0 595 842]{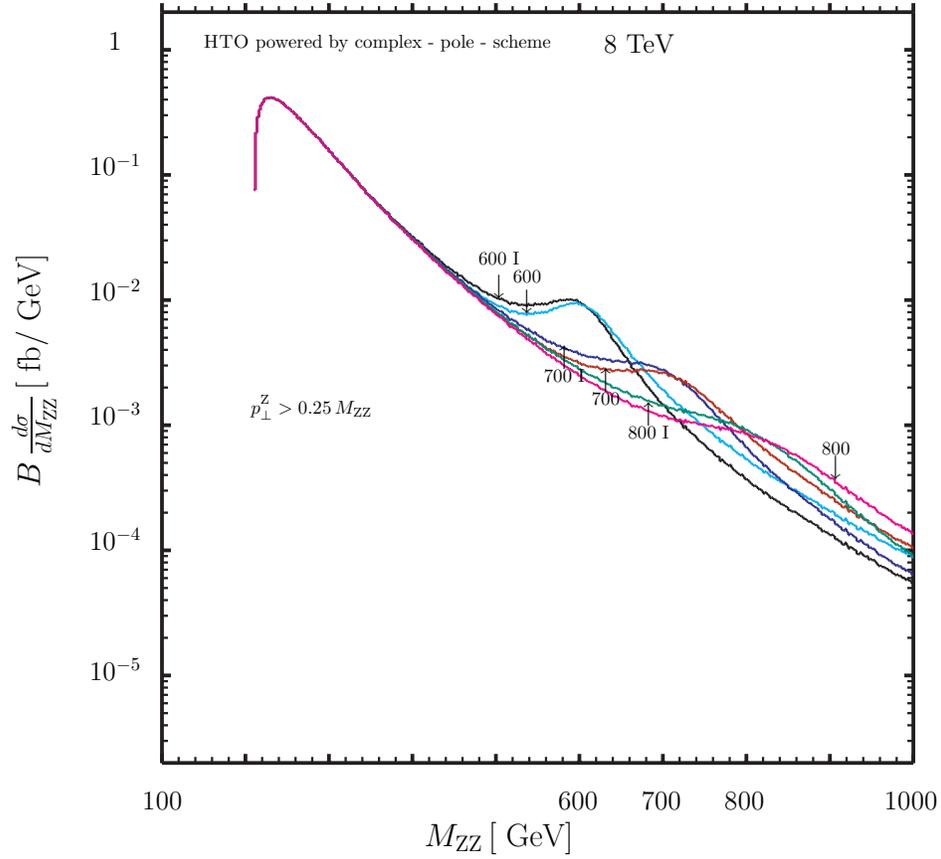}
  \vspace{-3.6cm}
  \caption{
LO Interference effects $[\%]$ in the $\PZ\PZ$ distribution for $\muh= 600,700,800\UGeV$.}
\label{fig:HTO_7}
\end{center}
\end{minipage}
\end{figure}

\begin{figure}
\vspace{-5cm}
\begin{minipage}{.9\textwidth}
\begin{center}
  \includegraphics[width=1.0\textwidth, bb = 0 0 595 842]{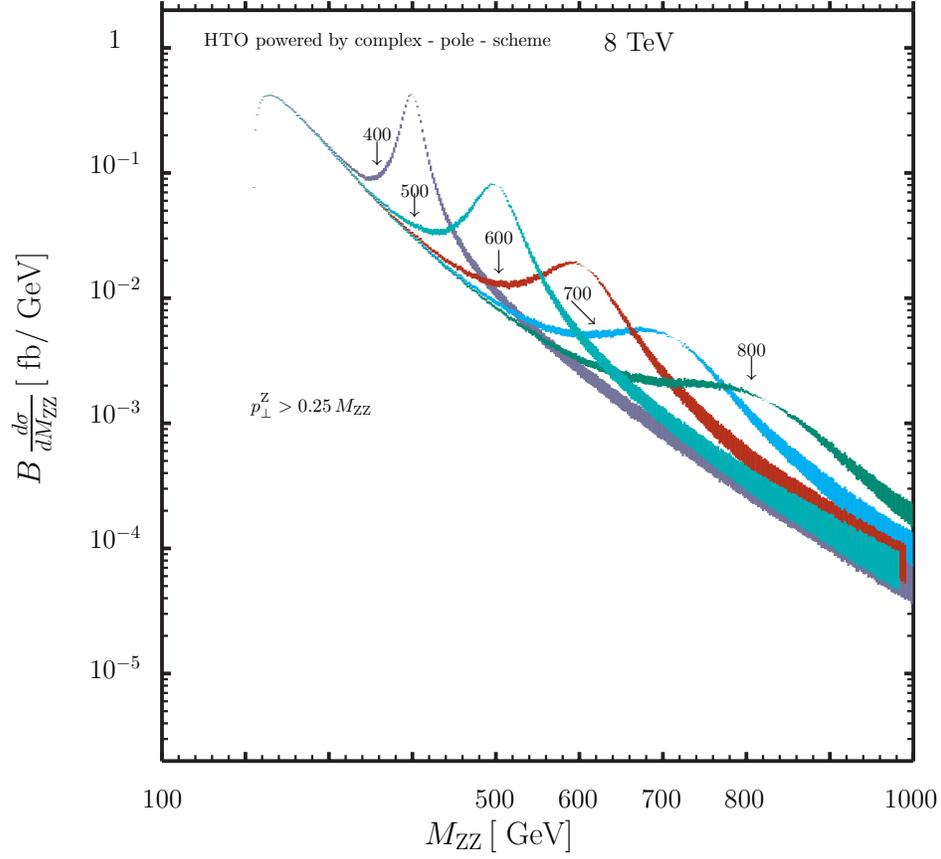}
  \vspace{-3.6cm}
  \caption{
Effective NNLO $\PZ\PZ$ invariant-mass distribution for $\muh= 400, 500, 600,700,800\UGeV$ 
including theoretical uncertainty.
$B = 4.36\,\cdot 10^{-3}$ represents the BR for both
$\PZ$ bosons to decay into $\Pe$ or $\PGm$.}
\label{fig:HTO_8}
\end{center}
\end{minipage}
\end{figure}

\begin{figure}
\vspace{-5cm}
\begin{minipage}{.9\textwidth}
\begin{center}
  \includegraphics[width=1.0\textwidth, bb = 0 0 595 842]{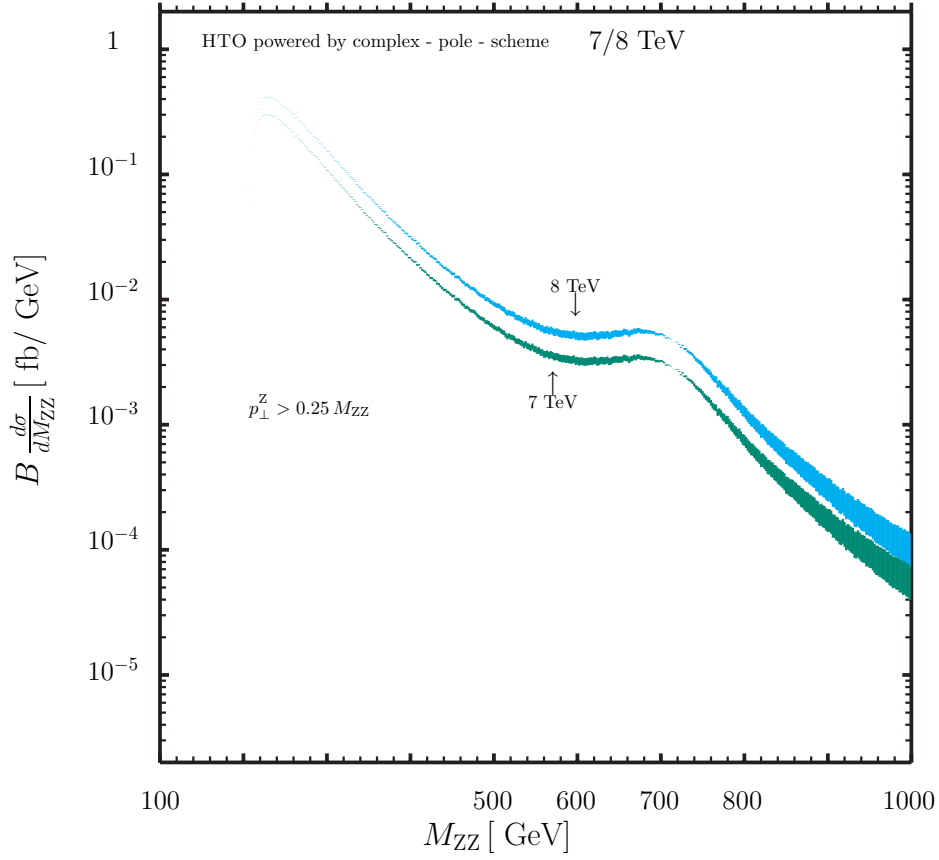}
  \vspace{-3.6cm}
  \caption{
Effective NNLO $\PZ\PZ$ invariant-mass distribution for $\muh= 700\UGeV$ including 
theoretical uncertainty and a comparison between $7\UTeV$ and $8\UTeV$.
$B = 4.36\,\cdot 10^{-3}$ represents the BR for both
$\PZ$ bosons to decay into $\Pe$ or $\PGm$.}
\label{fig:HTO_9}
\end{center}
\end{minipage}
\end{figure}

In \refF{fig:HTO_7} we present the LO interference effect for $\muh= 600, 700, 800\UGeV$.
In \refF{fig:HTO_8} we present the effective NNLO invariant mass distribution 
$\muh= 400, 500, 600, 700, 800\UGeV$, including our estimate of the theoretical uncertainty.
In \refF{fig:HTO_9} we show the effective NNLO $\PZ\PZ$ invariant-mass distribution 
for $\muh= 700\UGeV$ including theoretical uncertainty and a comparison between $7\UTeV$ and 
$8\UTeV$.

\begin{figure}
\vspace{-5cm}
\begin{minipage}{0.9\textwidth}
\begin{center}
  \includegraphics[width=1.0\textwidth, bb = 0 0 595 842]{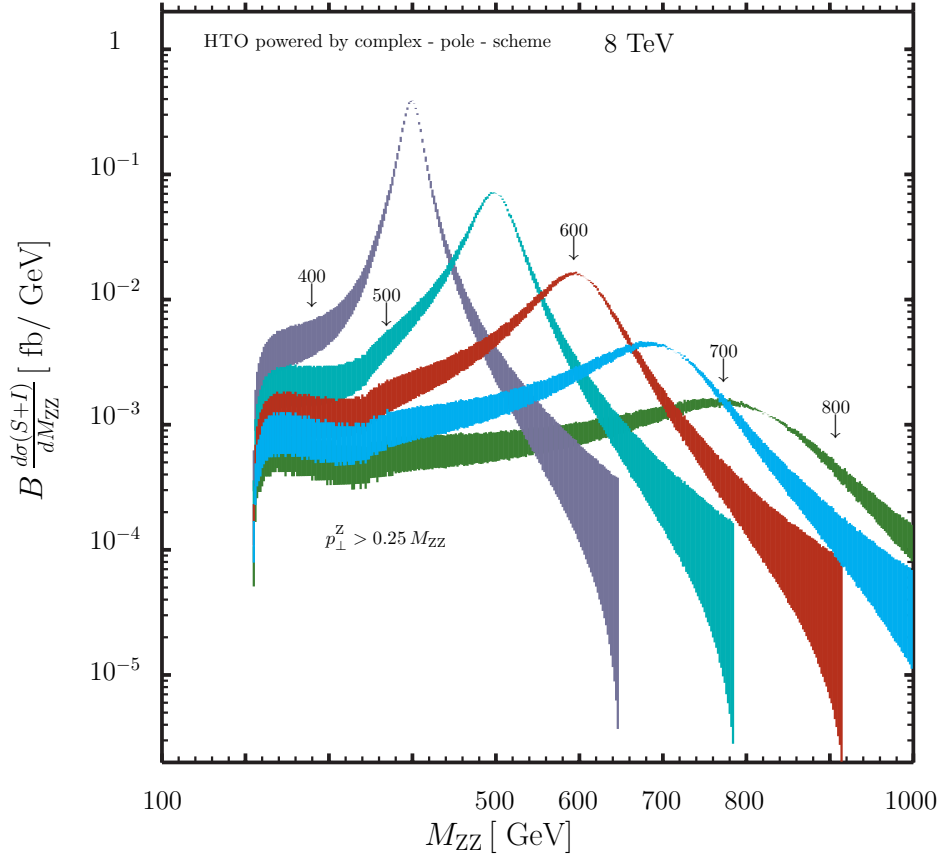}
  \vspace{-3.6cm}
  \caption{
Effective NNLO $\PZ\PZ$ invariant-mass distribution for $\muh= 400, 500, 600,700,800\UGeV$ 
including theoretical uncertainty. Only signal $\,+\,$ interference is plotted.
$B = 4.36\,\cdot 10^{-3}$ represents the BR for both
$\PZ$ bosons to decay into $\Pe$ or $\PGm$.}
\label{fig:HTO_11}
\end{center}
\end{minipage}
\end{figure}

In \refF{fig:HTO_11} we present the sum signal $\,+\,$ interference for
$\muh= 400, 500, 600, 700, 800\UGeV$, including our estimate of the theoretical
uncertainty. This quantity has no direct physical meaning but represents the
pseudo-observable preferred by the experimental Collaborations. 

\begin{figure}
\vspace{-5cm}
\begin{minipage}{0.9\textwidth}
\begin{center}
  \includegraphics[width=1.0\textwidth, bb = 0 0 595 842]{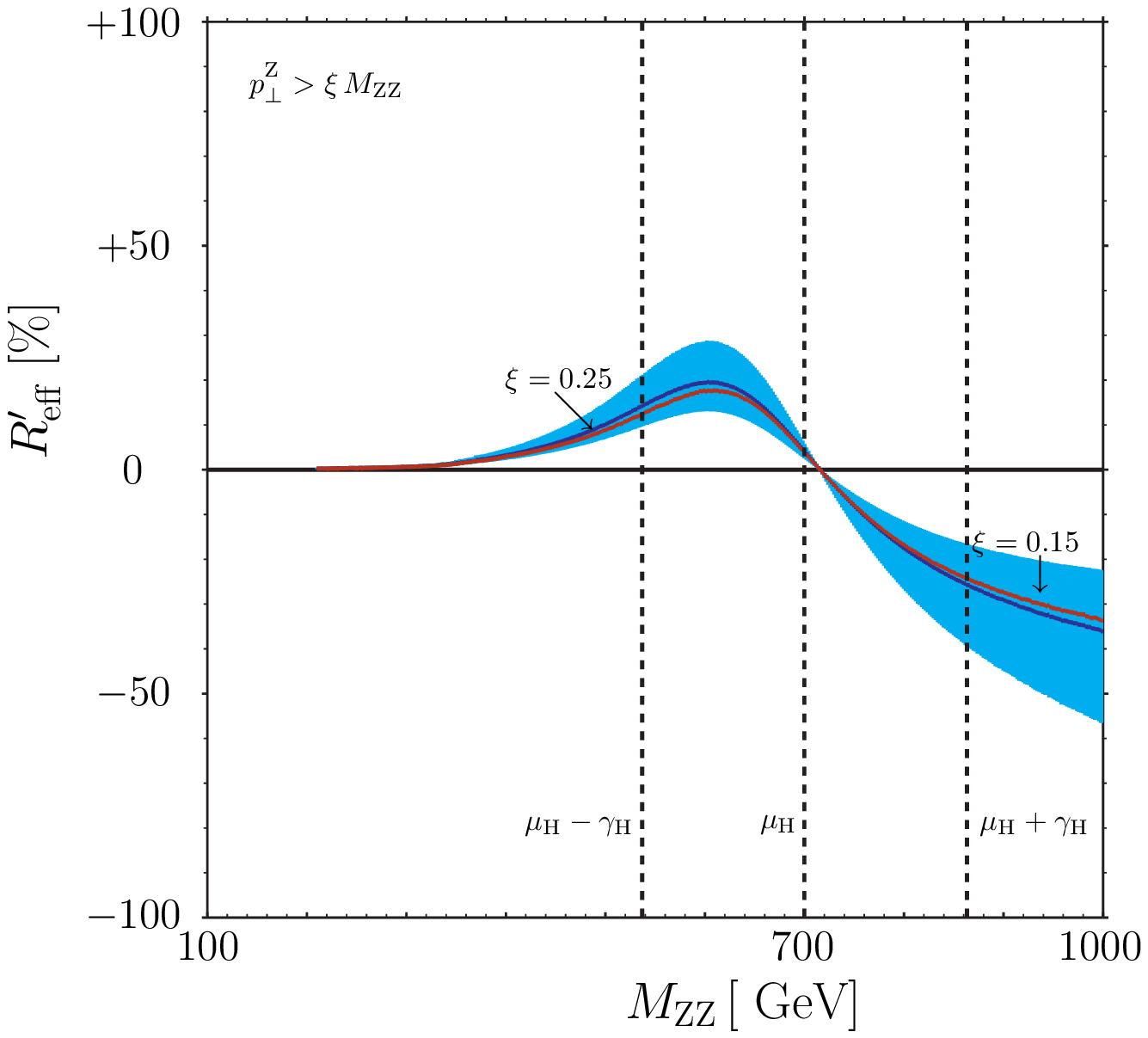}
  \vspace{-3.6cm}
  \caption{
Interference effects (see \eqn{percp}) in the $\PZ\PZ$ distribution for $\muh= 700\UGeV$
comparing $\pT^{\ssZ} > 0.25\,M_{\ssZZ}$ (blue) with $\pT^{\ssZ} > 0.15\,M_{\ssZZ}$ (red).}
\label{fig:HTO_10}
\end{center}
\end{minipage}
\end{figure}

In \refT{tab:HTO_4} we show the effect of the $\pT^{\ssZ}$ cut; from
$\pT^{\ssZ} > 0.25\,M_{\ssZZ}$ to $\pT^{\ssZ} > 0.15(0.05)\,M_{\ssZZ}$ the signal
is reduced by only $10\%(17\%)$ while the background is reduced by a factor $1.98(3.06)$. 
Finally in \refF{fig:HTO_10} we compare the interference effects with 
$\pT^{\ssZ} > 0.25\,M_{\ssZZ}$ and with $\pT^{\ssZZ} > 0.15\,M_{\ssZ}$; within the $\pm \gh$ 
window the change is negligible and becomes larger for lower or higher values of $M_{\ssZZ}$, 
as expected.

\begin{table}[ht]
\begin{center}
\caption[]{\label{tab:HTO_4}{Effect of the $\pT^{\ssZ}$ cut on the total LO cross-section 
(Signal, Background and Total multiplied by $B = 4.36\,\cdot 10^{-3}$, the BR for both
$\PZ$ bosons to decay into $\Pe$ or $\PGm$) for $2\,M_{\PZ} < M_{\ssZZ} < 1\UTeV$.}}
\vspace{0.2cm}
\begin{tabular}{rrrrr}
\hline 
                               & S\Ufb            & B\Ufb    & T\Ufb    & I/(S+B)[\%] \\
$\pT^{\ssZ} > 0.25\,M_{\ssZZ}$ & $1.091\,10^{-1}$ & $7.797$  & $7.971$  & $0.82$      \\
$\pT^{\ssZ} > 0.20\,M_{\ssZZ}$ & $1.163\,10^{-1}$ & $11.491$ & $11.683$ & $0.65$      \\
$\pT^{\ssZ} > 0.15\,M_{\ssZZ}$ & $1.216\,10^{-1}$ & $15.553$ & $15.760$ & $0.54$      \\
$\pT^{\ssZ} > 0.05\,M_{\ssZZ}$ & $1.274\,10^{-1}$ & $24.139$ & $24.366$ & $0.41$      \\
\hline
\end{tabular}
\end{center}
\end{table}
\subsection{Residual theoretical uncertainty}
In our results we have not included uncertainties coming from QCD scale variations and
from PDF$\,+\alphas$; due to the scaling of the LO result, these uncertainties are
coming from the numerator in the $K\,$-factor and are the typical NNLO 
uncertainties in gluon-gluon fusion~\cite{Dittmaier:2011ti}.
Also excluded is the residual electroweak uncertainty for the signal 
lineshape~\cite{Goria:2011wa}. To give an example of the complete set of theoretical
uncertainties we select $\muh = 700\UGeV$ and define maximum and half-maxima of the
signal lineshape; they are given by
\bq
S\lpar M_{\ssZZ} \rpar = \frac{d\sigma^{\ssS}}{d M_{\ssZZ}},
\quad
S\lpar M_1\rpar = \max\,S\lpar M_{\ssZZ} \rpar,
\quad
S\lpar M^{\pm}_{1/2}\rpar= \frac{1}{2}\,S\lpar M_1\rpar.
\eq
We find $M_1 = 701\UGeV$ and $M^-_{1/2}= 565\UGeV$, $M^+_{1/2}= 761\UGeV$. We define
theoretical uncertainties (THU) according to the following sources:
1) intrinsic, the full band between multiplicative and additive options, 2)
electroweak, due to THU on $\gh$ and $\Gamma_{\PH \to \PZ\PZ}(M)_{\ssZZ})$
and described in Sect.~7 of \Bref{Goria:2011wa} and 3) QCD scales.
THU arise from uncertainties in underlying theoretical paradigm. 
\begin{table}[hb]
\begin{center}
\caption[]{\label{tab:HTO_3}{Theoretical uncertainty on $R'_{\eff}= I/(S + B)$ for 
$\muh= 700\UGeV$.}}
\vspace{0.2cm}
\begin{tabular}{ccccc}
\hline 
$M_{\ssZZ}$ & $R'_{\eff}[\%]$  & intrinsic[\%]        & EW[\%]              &  QCD scales\\
$M_1$       & ${+}3.82$        & ${-}1.28\;{+}1.85$   & ${-}0.34\;{+}0.26$  &
${-}0.38\;{+}0.67$ \\
$M^-_{1/2}$ & ${+}17.19$       & ${-}5.59\;{+}8.02$   & ${-}0.34\;{+}0.44$  &
$< 0.1$             \\
$M^+_{1/2}$ & ${-}10.47$       & ${-}5.18\;{+}3.58$   & ${-}0.56\;{+}4.24$  &
${-}1.58\;{+}0.82$  \\
\hline
\end{tabular}
\end{center}
\end{table}
Results are shown in \refT{tab:HTO_3} for $R'_{\eff}= I/(S+B)$. THU on $M_1$ is tiny, intrinsic 
THU is large for the half-maxima, electroweak THU remains small in the window between the two 
half-maxima. As far as QCD scale variation is concerned we observe that $R'_{\eff}$ is very 
stable when varying $M_{\ssZZ}/4 < \muR,\muF < M_{\ssZZ}$. The percentage correction $R'_{\rm eff}$
is the scaling factor that one has to apply to her/his own calculation of $S + B$. What
it is meant in \refT{tab:HTO_3} is the following: the THU on $I$ induced by QCD scale
variation is between $5.39\%\,\bigl[S + B\bigr](M_{|ssZZ}/4)$ and 
$6.44\%\,\bigl[S + B\bigr](M_{\ssZZ})$ at $M_{\ssZZ} = M_1 = 701\UGeV$, etc  
The largest uncertainty in scale variation is due to the background which is only known
at LO; at $M_{\ssZZ} = M_1$ we observe a variation of $19\%$ in $K_{\ssD}\,S + B$  while the 
variation in $S + B$ is $89\%$. For this reason it would be difficult to work completely at LO.
\section{Conclusions \label{Sect_conc}}
In this paper we have addressed some issues concerning the inclusion of interference
effects in gluon-gluon fusion, especially for high values of the Higgs boson mass.

\begin{figure}
\vspace{-5cm}
\begin{minipage}{.9\textwidth}
\begin{center}
  \includegraphics[width=1.0\textwidth, bb = 0 0 595 842]{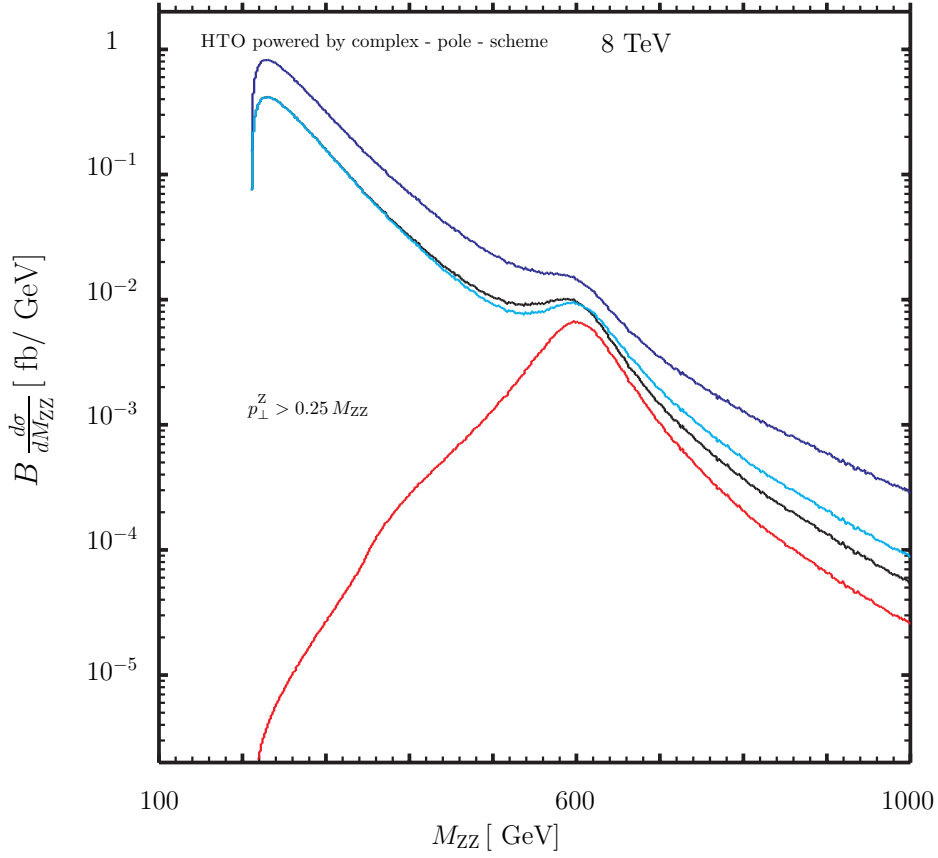}
  \vspace{-3.6cm}
  \caption{
The $\PZ\PZ$ invariant mass distribution in the OFFP-scheme of \Bref{Goria:2011wa} with
running QCD scales for $\muh= 600\UGeV$. $B = 4.36\,\cdot 10^{-3}$ represents the BR for both
$\PZ$ bosons to decay into $\Pe$ or $\PGm$.
The black line gives the full $\Pg\Pg \to \PZ\PZ$ process at LO; the cyan line gives signal plus
background (LO) neglecting interference while the blue line includes both $\Pg\Pg \to \PZ\PZ$ and
$\PAQq\PQq \to \PZ\PZ$ components (LO). The red line gives the LO signal.} 
\label{fig:HTO_1}
\end{center}
\end{minipage}
\end{figure}

\begin{figure}
\vspace{-5cm}
\begin{minipage}{.9\textwidth}
\begin{center}
  \includegraphics[width=1.0\textwidth, bb = 0 0 595 842]{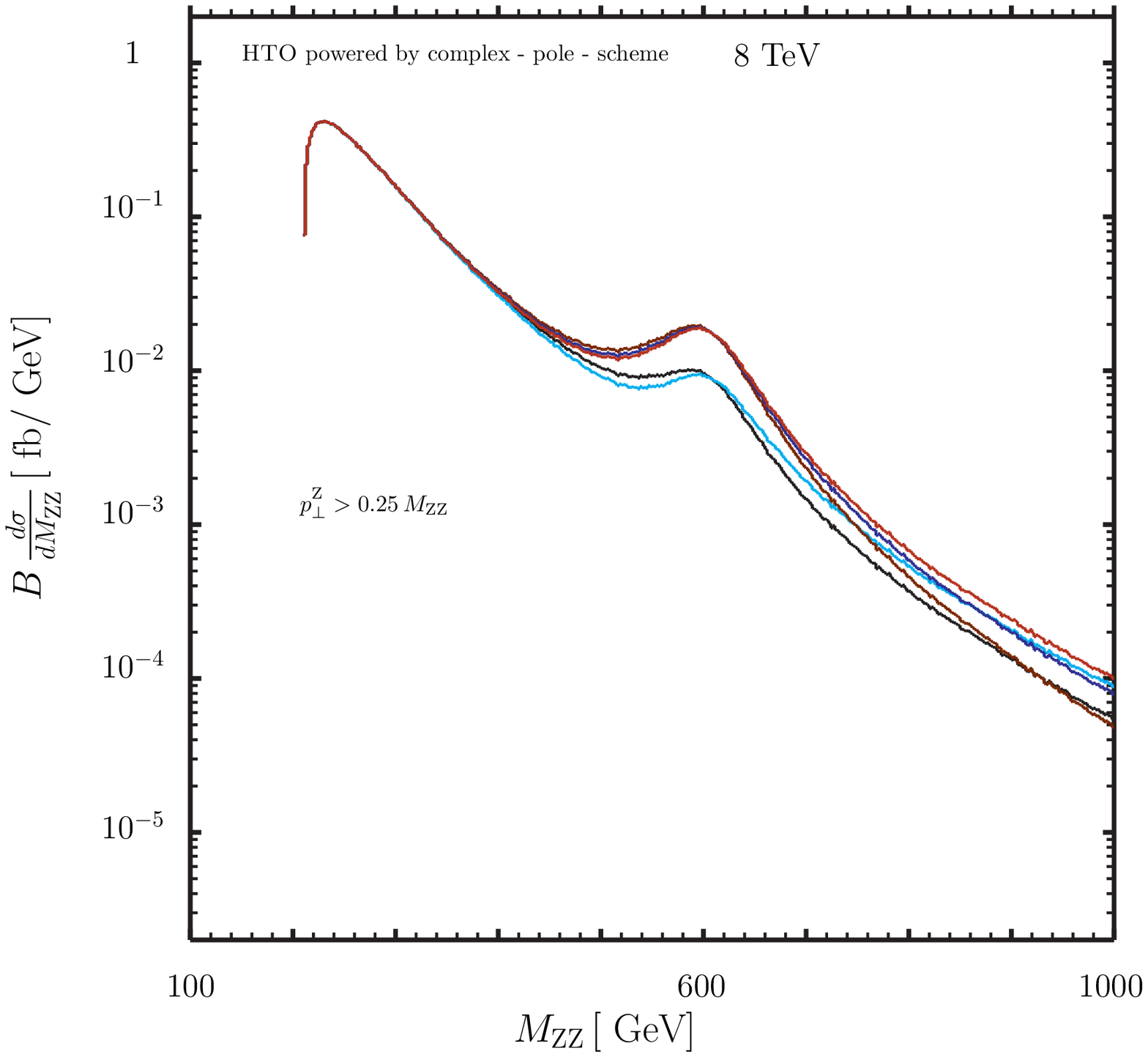}
  \vspace{-3.6cm}
  \caption{
The $\PZ\PZ$ invariant mass distribution in the OFFP-scheme of \Bref{Goria:2011wa} with
running QCD scales for $\muh= 600\UGeV$. $B = 4.36\,\cdot 10^{-3}$ represents the BR for both
$\PZ$ bosons to decay into $\Pe$ or $\PGm$.
The black line is the full LO $\Pg\Pg \to \PZ\PZ$
result, the brown line gives the multiplicative option of \eqn{Mopt}, the red line is the 
additive option of \eqn{Aopt} while the blue line is the intermediate option of \eqn{Iopt}. 
The cyan line gives signal plus background (LO) neglecting interference.}
\label{fig:HTO_2}
\end{center}
\end{minipage}
\end{figure}

\begin{figure}
\vspace{-5cm}
\begin{minipage}{.9\textwidth}
\begin{center}
  \includegraphics[width=1.0\textwidth, bb = 0 0 595 842]{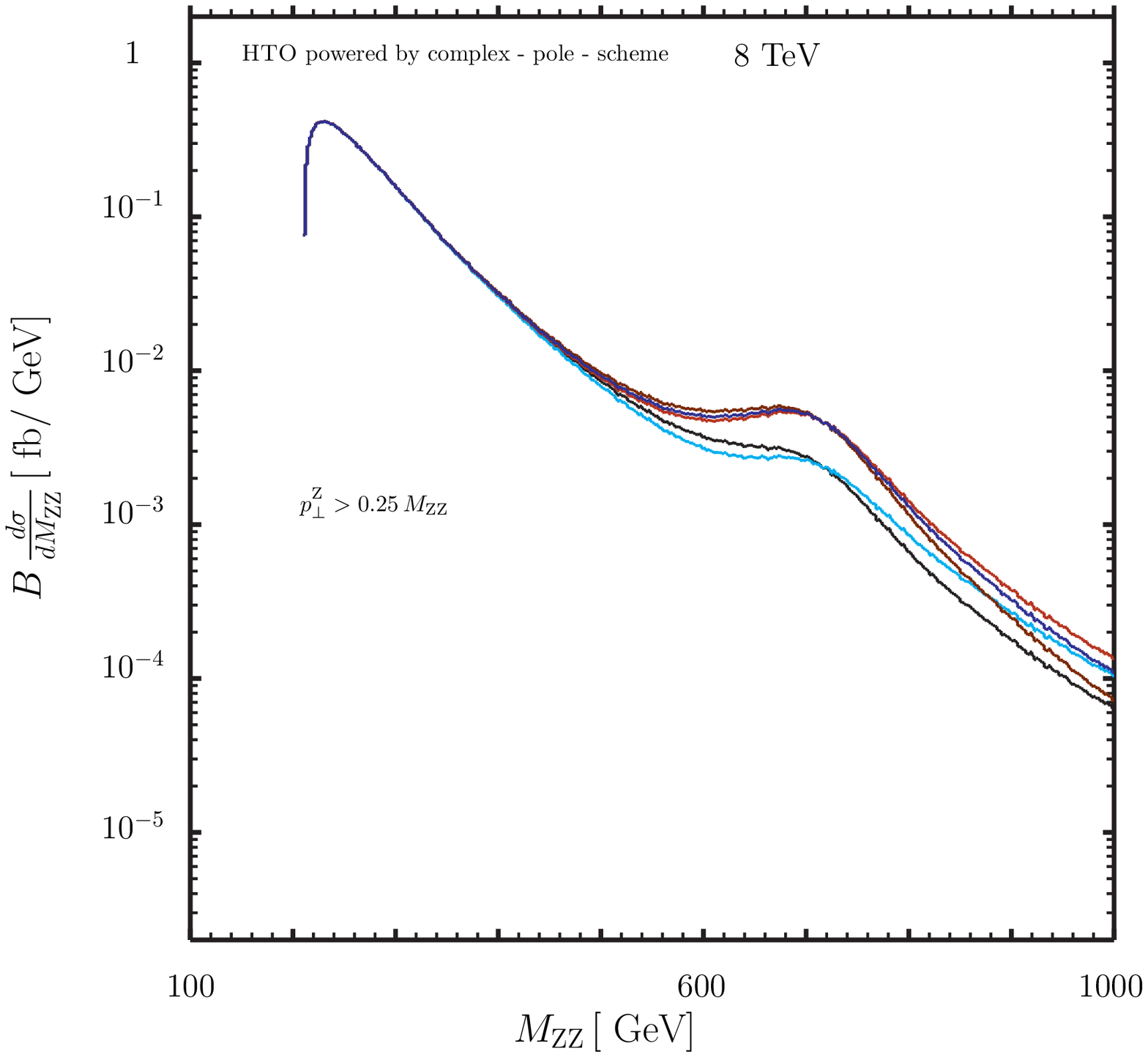}
  \vspace{-3.6cm}
  \caption{
The $\PZ\PZ$ invariant mass distribution in the OFFP-scheme of \Bref{Goria:2011wa} with
running QCD scales for $\muh= 700\UGeV$. $B = 4.36\,\cdot 10^{-3}$ represents the BR for both
$\PZ$ bosons to decay into $\Pe$ or $\PGm$.
The black line is the full LO $\Pg\Pg \to \PZ\PZ$
result, the brown line gives the multiplicative option of \eqn{Mopt}, the red line is the 
additive option  of \eqn{Aopt} while the blue line is the intermediate option of \eqn{Iopt}. 
The cyan line gives signal plus background (LO) neglecting interference.}
\label{fig:HTO_3}
\end{center}
\end{minipage}
\end{figure}

The results of \refF{fig:HTO_1} - \refF{fig:HTO_3} suggest the following
compromise for effectively including higher order effects in the interference
between Higgs and continuum contributions in $\Pg\Pg \to \PZ\PZ$. For the heavy Higgs
scenario, above the Higgs boson peak the {\em multiplicative} (or at least the {\em intermediate}) 
option is recommended while the {\em additive} (or the {\em intermediate}) one should be
preferred below the peak. However, one should also provide an estimate of the corresponding 
theoretical uncertainty. For this reason a conservative assessment of interference effects is
represented by a central value given by the {\em intermediate} option of \eqn{Iopt} with remaining
theoretical uncertainty given by the full band between the{\em additive} option of \eqn{Aopt} 
and the{\em multiplicative} option of \eqn{Mopt}.

For an inclusive quantity the effect of the interference, with or without the NNLO
$K\,$-factor for the signal, is almost negligible.
For distributions this is radically different and we have shown our results for the $\PZ\PZ$ 
invariant mass distribution: close to $M_{\ssZZ} = \muh$ the uncertainty is small
but becomes large in the rest of the search window $[\muh - \gh\,,\,\muh + \gh]$. 
The effect of the LO interference, w.r.t. LO $S + B$, reaches a maximum of ${+}16\%$ before the 
peak (\eg at $\muh=700\UGeV$) while our estimate of the scaled interference (always w.r.t. 
LO $S + B$) is $86\,{+}7\,{-}3\,\%$ in the same region, showing that NNLO signal effects 
are not negligible.
The estimate of the uncertainty is certainly a conservative one; however, it would be unsave 
to select less-conservative choices since we are not able to properly quantify the NLO corrections 
for the background. 
The percentage effect of the interference is slightly distorted when we go from LO to effective 
NNLO (at least in the {\em intermediate} option), however the global effect on the complete 
distribution is sizable, as seen in \refF{fig:HTO_3}.

In summary, we have discussed options to simulate NNLO corrections for the continuum interference
effects to the $\PZ\PZ\,$-signal for the Standard Model Higgs boson produced via gluon fusion. The
effects are large in the heavy Higgs-mass scenario, depending on the $\PZ\PZ$ invariant
mass. 

We can anticipate precisely what the likely criticism will be, therefore we must clearly 
state that LO kinematics is different from the NLO(NNLO) one and the $K\,$-factor will depend 
on aspects of kinematics that are not present in the LO background-interference, \eg contribution
to $\pT^{\ssZ}$ coming from emission of extra gluons \etc
Therefore, there is absolutely no guarantee that different distributions will not be distorted 
by the procedure and we will need to check the effect of the NLO effects on the interference 
for the full kinematic distributions.
However, one should remember that the band representing the theoretical uncertainty has no
statistical meaning, at most a flat prior to represent maximal uncertainty. Note that
the so-called {\em central value} also has no special meaning, although it is extremely useful 
for the experimental analysis where the MonteCarlo events are reweighted by using some analytical 
function. An alternative view considers the difference between multiplicative and additive options
as a systematic uncertainty resulting in overall distorsion of the shape. We can try to turn our
three {\em measures} of the lineshape into a continuous estimate in each bin; there is a
technique, called ``vertical morphing''~\cite{Conway:2011in}, that introduces a ``morphing''
parameter $f$ which is nominally zero and has some uncertainty. If we define
\bq
D^0 = \frac{d \sigma}{d M_{\ssZZ}}, \quad \mbox{option I},
\qquad
D^+ = \max_{\ssA,\ssM}\,D 
\quad
D^- = \min_{\ssA,\ssM}\,D, 
\eq
the simplest ``vertical morphing'' replaces
\bq
D^0 \to D^0 + \frac{f}{2}\,\lpar D^+ - D^-\rpar.
\eq
Of course, the whole idea depends on the choice of the distribution for $f$, usually 
Gaussian which is not necessarily our case; instead, one would prefer to maintain, as much as
possible, the LO cancellations around the peak\footnote{We gratefully acknowledge S.~Bolognesi for 
suggesting this alternative.}.
We would like to elaborate on this with the following heuristic argument: how does the lineshape
uncertainty band translate into uncertainty for the total cross-section?
We define two curves
\bq
C_{\ssM}\lpar \lambda\,,\,M_{\ssZZ}\rpar= \lambda\,D_{\ssI}\lpar M_{\ssZZ}\rpar + 
\lpar 1 - \lambda\rpar\,D_{\ssM}\lpar M_{\ssZZ}\rpar,
\qquad
C_{\ssA}\lpar \lambda\,,\,M_{\ssZZ}\rpar= \lambda\,D_{\ssI}\lpar M_{\ssZZ}\rpar + 
\lpar 1 - \lambda\rpar\,D_{\ssA}\lpar M_{\ssZZ}\rpar,
\eq
where $D_i, i=I,A,M$ is the lineshape according to $I,A$ and $M$ options. The parameter
$\lambda$, with $0 \le \lambda \le 1$, parametrizes how far we are from the central value
(we assume that $\lambda$ has a flat distribution).
The uncertainty on the total cross-section is obtained by integrating over $M_{\ssZZ}$, once
along $C_{\ssM}$ and a second time along $C_{\ssA}$; the difference gives the uncertainty,
maximal for $\lambda = 0$. The observation is not trivial since the two curves cross
shortly after the peak and one should not integrate over $\min\{C_{\ssA}\,,\,C_{\ssM}\}$
and over $\max\{C_{\ssA}\,,\,C_{\ssM}\}$.

It is worth noting that even for the signal alone, $ \Pg\Pg \to \PH \to \PV\PV$, the
current generation of events is done with something like {\sc POWHEG}~\cite{Alioli:2010xd} for 
the initial state kinematics plus {\sc PYTHIA}~\cite{Sjostrand:2006za,Sjostrand:2007gs} for 
the decay. The resulting events are rescaled in cross-section to NNLO for $\Pg\Pg \to \PH$ while
the branching ratios are rescaled according to 
{\sc Prophecy4f}~\cite{Bredenstein:2007ec,Prophecy4f}.
For the kinematics itself this means $\Pg\Pg \to \PH$ is NNLO while the $\PH$ decay 
has the LO model of {\sc PYTHIA}.

Further study is warranted of exact NLO background for all selected channels and
there is not much we can add at this stage; phases in NLO(NNLO) corrections to the signal
are not available as well as NLO corrections to the background. 
The typical example that we have in mind is the process $\Pp\Pp \to \PGg\PGg\PX$, as described
in \Bref{Catani:2011qz} for the full NNLO QCD corrections; in the $\Pg\Pg\,$-channel the 
box contribution was computed in \Bref{Dicus:1987fk} and the next-order gluonic corrections 
in \Bref{Bern:2002jx}. This is a case where all contributions for the $\Pg\Pg\,$-channel
have been included in a fully-consistent manner.

The options that have been introduced in the literature and that we have summarized and extended 
here can only give us a rough approximation of the true result. 
However, this solution improves upon the previous estimate of $150\,\muh^3[\%]$ ($\muh$ in \UTeV)
made without any reference to the existing LO calculation of the interference.

The calculation described in this work refers to a SM Higgs boson; however, ATLAS and CMS
quantify their analysis in terms of 
\bq
\mbox{Hypothesis}(\mu) = \mu\,\sigma(S) + \sigma(B)
\label{Hy0}
\eq
where interference is only included in the uncertainty; $\mu= 0$ is the null 
hypothesis, $\mu \not= 0$ is the alternate hypothesis (Neyman - Pearson lemma).
Another issue, currently under discussion, is the following: should one define the
cross-section for events containing a Higgs boson in terms of $sigma(S) + \sigma(I)$
with $\sigma(B)$ defining the null hypothesis? The obvious criticism is that 
signal $\,+\,$ interference is not positive defined.
Therefore, it is not completely clear how the hypothesis should be modified to include
the interference~\cite{Mistlberger:2012rs}, a possibility being~\cite{Bab}
\bq
\mbox{Hypothesis}(\mu) = \mu\,\sigma(S) + \sqrt{\mu}\,\sigma(I) + \sigma(B),
\label{Hy1}
\eq
which is reasonably good for $\mu \approx 1$, \ie for analyzing the SM hypothesis. In
general it would be better to start with an effective Lagrangian containing anomalous 
couplings~\cite{Anastasiou:2011pi} $\{\lambda\}$ (such that $\lambda_i = 0$ is the SM) and to 
define
\bq
\mbox{Hypothesis}(\{\lambda\}) = \sigma(S,\{\lambda\}) + \sigma(I,\{\lambda\}) + 
                                 \sigma(B,\{\lambda\}),
\label{AC}
\eq
which will respect, among other things, the unitarity cancellations requested by \eqn{asym},
\ie
\bq
\mbox{Hypothesis}(\mu)\bmid_{M_{\ssZZ} \to \infty} = 0,
\label{unit}
\eq
which proves that, in any consistent theory (not only the SM), the background {\em knows}
the signal (at least asymptotically). Clearly, \eqn{unit} imposes $\mu = 1$ in 
\eqn{Hy1} if $\sigma(S)$ \etc are the SM cross-sections.
A possibility in implementing \eqn{AC} is to use the 
Buchm\"uller - Wyler basis~\cite{Buchmuller:1985jz}.

If one does not have a clear idea of what the BSM signal is, it is difficult to optimize 
an analysis (Neyman - Pearson lemma cannot help). Alternative strategies require that, in 
absence of a signal, a C.L. limit is set on $\muh$. Based on the observation that a ultra-heavy 
Higgs boson does not lead to a pronounced peak structure, Baur and 
Glover~\cite{Baur:1990af,Baur:1990mr} made an alternative proposal for the null hypothesis: here 
the background corresponds to the minimum of $S + I + B$ for all Higgs masses. More precisely, 
in $\PZ\PZ\,$-production, they define
\bq
S_{\max} = \max\,S\lpar m_0 \rpar,
\qquad
S^2(m_0) = \frac{\bigl[ N\lpar \muh\,,\,m_0\rpar - N\lpar \muph\,,\,m_0\rpar\bigr]^2}{N\lpar \muph\,,\,m_0\rpar},
\eq
where $M_{\ssZZ} > m_0$ and $N\lpar \muh\,,\,m_0\rpar$ is the number of signal events for
a Higgs boson of mass $\muh$. Therefore, $S_{\max}$ gives a quantitative measure of how
well the $\muh$ hypothesis can be discriminated from the $\muph$ hypothesis. The background 
corresponds to $\muph = 0$ but we could have $\muh= 125\UGeV$ as well.
The strength of the analysis should be in terms of the C.L. at which the $\muh$ hypothesis 
can be excluded. 
Note that model-independent searches have been addressed in \Bref{Abbott:2000fb,Aktas:2004pz}
and recommendations have been made in \Bref{Kraml:2012sg}.

If the signal for a light Higgs boson will be confirmed at LHC then any heavy scalar
boson must be beyond-Standard-Model (BSM). As a prototype we can take a THD model; from the 
perspective of searching for the heavy partner(s) the light one is assimilable to the
background (\ie the two resonances do not interfere). Of course, the signal has to be rescaled 
according to the BSM model (couplings of the heavy partner(s), $\PH, \PA, \PH^{\pm}$, to 
fermions and gauge bosons) but we have not attempted a detailed analysis, \eg taking
into account the mass difference among the heavy bosons, related to the breaking of the
custodial $SU(2)$ symmetry. Other possibilities include a heavy scalar singlet with a large 
vacuum expectation value that can evade the potential instability of the SM electroweak 
vacuum~\cite{EliasMiro:2012ay}. The qualitative aspects of this work will not change.

Finally, the interference effects for a light SM Higgs boson will follow the same pattern 
described in this work; the final state will contain four fermions (below the $\PZ\PZ$ or 
$\PW\PW$ thresholds) but the background/interference will always be at LO and we will be 
missing large $K\,$-factors. Numerics will change but our recipe for estimating effective NNLO 
signal $\,+\,$ background $\,+\,$ interference and the corresponding theoretical uncertainty will
remain the same. Only progress, \ie new more accurate calculations will be able to produce 
more accurate estimates for the theoretical uncertainty.

\Acknowledgments
We gratefully acknowledge several important discussions with S.~Bolognesi, D.~de~Florian,
M.~Duehrssen, E.W.~Glover, M.~Mangano, C.~Mariotti, F.~Petriello and R.~Tanaka. 
We gratefully acknowledge the invaluable assistance of A.~Maier. 
This work has been performed within the Higgs Cross-Section Working Group\\
{\tt https://twiki.cern.ch/twiki/bin/view/LHCPhysics/CrossSections}.

\clearpage

\begin{table}[h]
\begin{center}
\caption[]{\label{tab:HTO_2}{Interference effect (percentage), $R'_{\eff}$ for
$\muh= 700\UGeV$ with corresponding theoretical uncertainty. 
Here $\pT^{\ssZ} > 0.25\,M_{\ssZZ}$ and $2\,M_{\PZ} < M_{\ssZZ} < 1\UTeV$.}}
\vspace{0.2cm}
\begin{tabular}{rrrr}
\hline 
Bin[\UGeV] & $R'_{\eff}[\%]$ & minus error[\%] & plus error[\%] \\
$210 - 212$ & $0.24$ & ${-}0.07$ & ${+}0.11$\\
$230 - 232$ & $0.31$ & ${-}0.09$ & ${+}0.13$ \\
$250 - 252$ & $0.38$ & ${-}0.12$ & ${+}0.17$ \\
$270 - 272$ & $0.47$ & ${-}0.14$ & ${+}0.20$ \\
$290 - 292$ & $0.56$ & ${-}0.17$ & ${+}0.24$ \\
$310 - 312$ & $0.73$ & ${-}0.22$ & ${+}0.32$ \\
$330 - 332$ & $0.96$ & ${-}0.29$ & ${+}0.41$ \\
$350 - 352$ & $1.47$ & ${-}0.44$ & ${+}0.63$ \\
$370 - 372$ & $2.21$ & ${-}0.67$ & ${+}0.96$ \\
$390 - 392$ & $2.82$ & ${-}0.87$ & ${+}1.25$ \\
$410 - 412$ & $3.67$ & ${-}1.14$ & ${+}1.64$ \\
$430 - 432$ & $4.72$ & ${-}1.48$ & ${+}2.13$ \\
$450 - 452$ & $6.00$ & ${-}1.90$ & ${+}2.74$ \\
$470 - 472$ & $7.49$ & ${-}2.37$ & ${+}3.41$ \\
$490 - 492$ & $9.24$ & ${-}2.95$ & ${+}4.24$ \\
$510 - 512$ & $11.30$ & ${-}3.63$ & ${+}5.21$ \\
$530 - 532$ & $13.49$ & ${-}4.36$ & ${+}6.27$ \\
$550 - 552$ & $15.87$ & ${-}5.14$ & ${+}7.37$ \\
$570 - 572$ & $17.75$ & ${-}5.78$ & ${+}8.30$ \\
$590 - 592$ & $19.12$ & ${-}6.25$ & ${+}8.97$ \\
$610 - 612$ & $19.43$ & ${-}6.39$ & ${+}9.19$ \\
$630 - 632$ & $18.16$ & ${-}5.99$ & ${+}8.62$ \\
$650 - 652$ & $15.39$ & ${-}5.11$ & ${+}7.36$ \\
$670 - 672$ & $11.29$ & ${-}3.77$ & ${+}5.42$ \\
$690 - 692$ & $6.43$ & ${-}2.15$ & ${+}3.10$ \\
$710 - 712$ & $1.28$ & ${-}0.43$ & ${+}0.62$ \\
$730 - 732$ & ${-}3.72$ & ${-}1.82$ & ${+}1.26$ \\
$750 - 752$ & ${-}8.34$ & ${-}4.11$ & ${+}2.84$ \\
$770 - 772$ & ${-}12.43$ & ${-}6.18$ & ${+}4.26$ \\
$790 - 792$ & ${-}16.05$ & ${-}8.06$ & ${+}5.55$ \\
$810 - 812$ & ${-}19.21$ & ${-}9.73$ & ${+}6.69$ \\
$830 - 832$ & ${-}21.98$ & ${-}11.19$ & ${+}7.69$ \\
$850 - 852$ & ${-}24.36$ & ${-}12.54$ & ${+}8.58$ \\
$870 - 872$ & ${-}26.52$ & ${-}13.88$ & ${+}9.45$ \\
$890 - 892$ & ${-}28.37$ & ${-}14.98$ & ${+}10.18$ \\
$910 - 912$ & ${-}30.06$ & ${-}16.07$ & ${+}10.88$ \\
$930 - 932$ & ${-}31.63$ & ${-}17.18$ & ${+}11.57$ \\
$950 - 952$ & ${-}33.01$ & ${-}18.10$ & ${+}12.16$ \\
$970 - 972$ & ${-}34.28$ & ${-}19.17$ & ${+}12.79$ \\
$990 - 992$ & ${-}35.50$ & ${-}20.11$ & ${+}13.36$ \\
\hline
\end{tabular}
\end{center}
\end{table}

\clearpage

\clearpage

\clearpage

\clearpage

\clearpage

\clearpage

\clearpage

\clearpage

\clearpage

\clearpage

\clearpage

\bibliographystyle{atlasnote}
\bibliography{HLS}{}

\end{document}